# Collaboration Networks in The Music Industry

**Pascal Budner**

University of Cologne

Information Systems

`pbudner@smail.uni-koeln.de`

**Jörn Grahl**

Digital Transformation and Analytics

Information Systems

University of Cologne

`grahl@wiso.uni-koeln.de`

http://www.digital.uni-koeln.de

http://www.album-network.uni-koeln.de/

**Abstract**

Recording an album brings singers, producers, musicians, audio engineers, and many other professions together. We know from the press that a few "super"-producers work with many artists. But how does the large-scale social structure of the music industry look like? What is the social network behind the finest albums of all time? In this paper we studied the large-scale structure of music collaborations using the tools of network science. We considered all albums in Rolling Stone Magazine's list of '500 Greatest Albums of All Time' and the '1001 Albums You Must Hear Before You Die' by Robert Dimery. We found that the existing research on collaboration networks is corroborated by the particular collaboration network in the music industry. Furthermore, it has been found that the most important professions of the music industry in terms of connectivity were main artists and engineers.

## 1. Introduction

According to Hara et al. (2003), collaboration can be defined as a social process of at least two or more partners working together to achieve a common goal and share their knowledge. Since each partner in a collaboration has different skills and knowledge, collaboration facilitates the solving of complex problems by joining a variety of expertise (Mattessich and Monsey, 1992).

The corresponding partners in a collaboration can either be organizations (Mattessich and Monsey, 1992) and/or individuals (Schrage, 1995).

Throughout the literature, several studies already investigated the construct collaboration using bibliometric methods (Subramanyam, 1983), qualitative methods (Hara et al., 2003), quantitative methods (Birnholtz, 2006), and network science (Barabási et al., 2002). In particular, network science uses collaboration networks to analyze the structure and dynamics of collaborations. Collaboration networks represent individuals or organizations as nodes and the collaborations itself as links between nodes.

Recent studies evidenced the existence of small-world and scale-free network topologies in collaboration networks (Barabási et al., 2002; Newman, 2001). Small-world networks are networks with a small average path length and a high local clustering (Albert and Barabási, 2002). The scale-free topology means that the distribution of the number of collaborations/links each node has follows a power-law, which means that there are many nodes with only a few links but also a few nodes with a large number of links (Barabasi, 2016). Additionally, other studies revealed preferential attachment, which means that well-connected nodes are more likely to establish new links (Barabasi, 1999).

Investigations concerning the detailed structures of collaboration networks focusing on the composition of the giant component, which is the biggest connected component of a network that contains the largest fraction of nodes of the network (Newman, 2010), have not been conducted in the existing literature. This paper fills the lack of knowledge by investigating the impact of different collaborator groups on the size of the giant component of a collaboration network in the context of the music industry.

Since collaboration in the music industry is common practice and recent network science studies already investigated the particular collaboration networks (Gleiser and Danon, 2003; Smith, 2006), this paper also performs the analysis on a genuine network of music collaborations.

## 2. Methodological Approach

The principal source of data for this paper is Discogs, which is a user-built digital library for discography and contains more than 7,700,000 recordings and 4,700,000 artists[1]. A set of representative albums has been selected from the list 'Rolling Stone's 500 Greatest Albums of

---

[1] See: https://www.discogs.com/about (Last visited 10/10/2016)

All Time'[2] and from the book '1001 Albums You Must Hear Before You Die' by Robert Dimery[3]. Both lists are editorially selected lists that claim to name the most important and influential albums of all time. Since Discogs provides an API to simplify the access to the data and offers a link list for the albums of both editorial lists, a crawler has been programmed that accesses the master releases of each album and fetches information about the album and the corresponding collaborators.

The gathered data consisting of albums and collaborators can be represented by a bipartite network. A bipartite network consists of two types of nodes: the actual nodes or collaborators and the corresponding nodes representing the groups or albums to which the actual nodes belong (Newman, 2010).

Finally, both one-mode projections of the bipartite network are analyzed. First, general topologies are analyzed. Afterwards, both one-mode projections are analyzed regarding the impact of the different collaborator groups on the giant component. The impact is investigated by excluding the links of each collaborator group and calculating the reduced size of the giant component. The lower the size of the giant component is, the higher is the impact of the specific collaborator group. The Python library NetworkX is used to analyze the statistics of the giant component.

3. Network Preparation

**Data Preparation.** The data for the collaboration network has been crawled via a web crawler resulting in 1,175 albums released from 1955 to 2010, 9,604 collaborators, 14,108 collaborator-to-album associations, and 16,648 roles of collaborators in an album have been fetched. This leads to a rate of 12.01 collaborators per album and 1.18 roles per collaborator in an album, which indicates that collaborators exist that serve more than one role in an album collaboration. Since the total number of albums is 1,175, both sources for the albums of this network have an overlap of 326 albums.

Moreover, the gathered dataset has been cleaned, which decreased the total number of unique roles from 2,216 to 257. For instance, the names of the collaborator roles have been cleaned by applying the following four modifications: (1) all roles have been lowercased, (2) all hyphens ("-") have been replaced by spaces, (3) the word "by" has been cut out, and (4) specialized roles have been cut out to create more comprehensive collaborator roles and a more common labeling. For instance, the original roles "Written by", "Written [all songs]", and "Written-By [co-written]" are finally assigned to the more generic role "written".

---

[2] See: http://www.rollingstone.com/music/lists/500-greatest-albums-of-all-time-20120531

[3] See: https://www.amazon.de/1001-Albums-Must-Hear-Before/dp/1844037355

**Network Creation.** Based on the cleaned data a bipartite network has been created. The first node group of the bipartite network are the 1,175 albums. The other group are the 9,604 collaborators. An album and a collaborator are considered as connected if a collaborator-to-album association is set, which means that a collaborator participated in the collaboration of an album. The resulting bipartite network can be projected onto the albums and onto the collaborators. A projection onto the collaborators leads to the typical social network (Newman, 2010), whereas a projection onto the albums leads to an album network that represents the teams working on an mutual album.

Performing a one-mode projection onto the albums using the number of mutual collaborations as the weight of the edges leads to the album network[4] that is analyzed in the next section. The nodes in this network represent albums/teams consisting of at least one collaborator. Two nodes are connected if they share a common collaborator.

Furthermore, a typical social network is created performing a one-mode projection onto the collaborators using the number of mutual collaborations as the weight of the edges. The resulting network is visualized in Figure 1. It is seen that several small components exist, which are not connected to the giant component in the center. Furthermore, the deep dark areas of the visualized network imply cliques. This network is also analyzed in the next section.

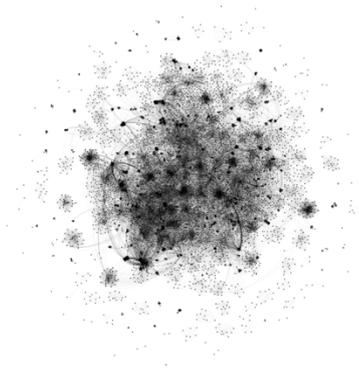

Figure 1: The visualized social network. Collaborators are represented by nodes and interconnected if they share a mutual album they have worked on.

Besides the base networks, 258 role-specific networks for each one-mode projection have been created for each omitted collaborator role using the following rule: An existing link from the base network is cut from the role-specific network if the link is completely dependent on the omitted role. For instance, if a link has a weight of two in the base network (e.g. one link with the role "producer" and one link with the role "engineer") and the role-specific network without producers is going to be created, the link is not cut but its weight is decreased by one.

---

[4] An interactive visualization of this network can be found at http://www.album-network.uni-koeln.de/

## 4. Network Analysis

Since both one-mode projections of the collaboration networks provide different views on the dynamics and structures of the collaboration network, both projection networks are analyzed in detail in this section. For instance, the nodes in the album network represent albums or teams consisting of at least one collaborator. Thus, this one-mode projection indicates the dynamics between teams in the collaboration network. The collaborator roles with the biggest impact on the connectivity of the album network are those that participate in different teams and, thus, establish the most bridges between the teams.

On the contrary, nodes in the social network represent the actual collaborators. Hence, this one-mode projection indicates social structures and dynamics. Each team or album in the album network is represented by a clique in the social network, since all collaborators of an album are interconnected to the other collaborators of the album. The impact of the collaborators roles on the social network is affected by two components: (1) the number of collaborators representing the particular role and (2) the number of bridges established by the collaborator.

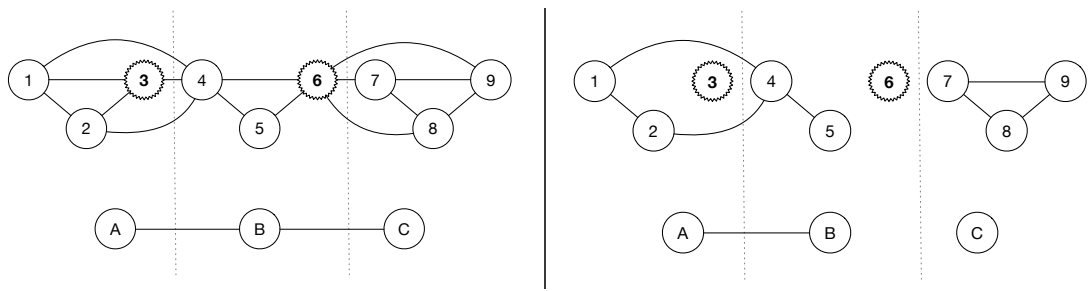

Figure 2: Different effects of omitting the connections of one collaborator role onto both one-mode projections of the collaboration network. The networks on the left represent the base network where no connections have been omitted. The upper networks represent the social network, whereas the lower network the album network.

**Basic Statistics.** The basic statistics of both one-mode projections are listed in Table 1. It is seen that the network density is small in both networks. Furthermore, 109 individual connected components exist, which is caused by the fact that the underlying data for the networks is only an extract from the real world. Thus, some of the crawled albums are not connected to the giant component of the network. Moreover, it can be seen that the average weighted degree is higher than the average degree, which leads to the conclusion that some links between nodes are caused by multiple mutual collaborations. The average path length is 3.4 or 4.0 and the diameter is 10 or 11. Considering that the networks consist of 1,175 or 9,604 nodes, it is concluded that the networks are small, since every connected node can be reached in average by three or four links.

| Characteristic | Album Network | Social Network |
|---|---|---|
| Number of nodes | 1,175 | 9,604 |
| Number of links | 8,798 | 126,153 |
| Density | 0.013 | 0.003 |
| Diameter | 10 | 11 |
| Average degree | 14.975 | 26.271 |
| Average weighted degree | 20.936 | 27.831 |
| Average clustering coefficient | 0.515 | 0.9 |
| Average path length | 3.402 | 4.003 |
| Number of connected components | 109 | 109 |

Table 1: Graph specific statistics of both one-mode projection networks.

Figure 3 visualizes the degree distributions of both networks. It can be seen that the degree distribution is for both networks right-skewed ($skewness_a = 1.6$, $skewness_s = 4.18$). Hence, even though there is a huge number of nodes with a low degree there are still nodes with a very high degree (hubs). These hubs are represented by the plateau in the log-log-scaled graph for the album network at y-value 0.10% and for the social network at y-value 0.01%. The top five of the biggest hubs in the networks are listed in Table 2.

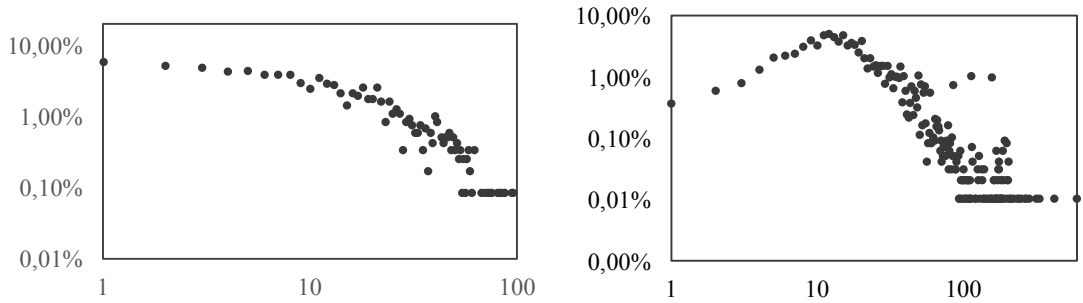

Figure 3: The degree distribution both networks in log-log scaled graphs. The degree $k$ is on the x-axis and the fraction of nodes $p_k$ having a degree of $k$ is represented on the y-axis. The graph on the left represents the album network and the graph on the right the social network.

| | **Album Network** | | | **Social Network** | | |
|---|---|---|---|---|---|---|
| # | Degree | Artist | Album | Degree | Name | Primary Role |
| 1 | 95 | Radiohead | In Rainbows | 549 | Bob Ludwig | Mastered |
| 2 | 94 | Bob Dylan | Live 1966 | 416 | Howie Weinberg | Mastered |
| 3 | 87 | John Lennon | Imagine | 331 | Anton Corbijn | Photography |
| 4 | 86 | Crowded House | Woodface | 314 | Greg Calbi | Mastered |
| 5 | 83 | Nirvana | Nevermind | 280 | Ted Jensen | Mastered |

Table 2: Top 10 of the largest hubs in both networks.

**Structure of the Giant Component**. According to Newman (2010), the giant component grows its size in proportion to the size *n* of the network. This dynamic growth has been analyzed in this paper using the release dates of the albums as the temporal progress of the network. Figure 4 visualizes the progress on the size of the giant component in comparison to the total number of nodes in the both networks. It is seen that the size of the giant component steadily grows in proportion to the size *n* of the network. This correlation is also supported by calculating the correlation coefficient between both measures. The resulting correlation coefficients $r_a = 0.9997$ and $r_s = 0.9998$ underline a strong positive correlation between the size of the network and the size of the giant component. The giant component in the final album network in 2010 has a size of 1,058 nodes, which is 90% of all nodes of the network. In contrary, the giant component in the social network in 2010 has a size of 9,072 nodes, which is 94.46% of all nodes in the network.

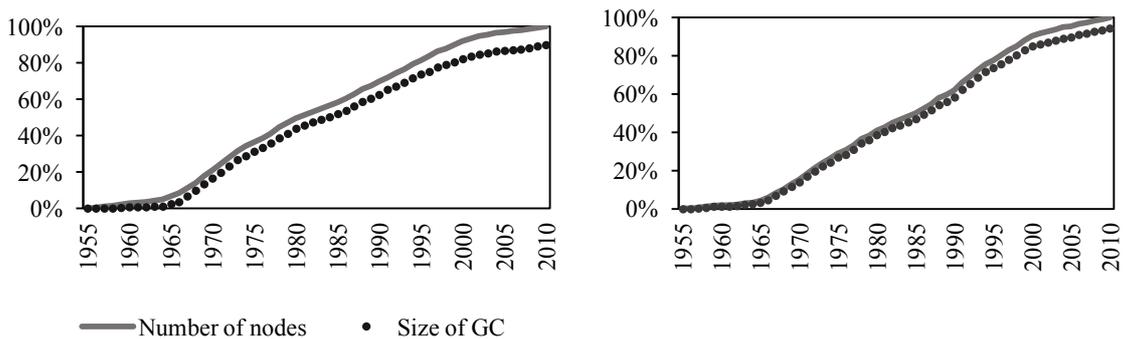

Figure 4: Dynamic growth of the giant component with the album release dates as the temporal progress. The left graph represents the album network, whereas the right graph represents the social network.

Since this paper investigates whether differences in the impact of specific collaborator roles on the connectivity of the network exist, the size of the giant component has been analyzed for each single omitted collaborator. For this purpose, the role-specific networks have been analyzed regarding their size of the giant component. Table 3 illustrates descriptive statistics of the role-specific networks based on the album network that lead to the following two findings:

First, it can be seen that the minimum size of the giant component still contains 86% of all nodes, even though the maximum number of omitted edges is high with 30.1% of all edges. Thus, the maximum impact of a single collaborator role on the connectivity of the giant component in the album network is only a decrease of four percentage points.

Second, it is seen that the number of omitted edges highly scatters, since the standard deviation has a high value of 203.08. Thus, it has to be considered whether the influence of a collaborator group depends only on the total number of omitted edges or also on other factors such as the importance of the omitted edges.

| Characteristic | Min | Max | Mean | $\sigma$ |
|---|---|---|---|---|
| Number of omitted edges | 0 | 2,649 | 37.9 | 203.08 |
| Number of edges | 6,149 | 8,798 | 8,760 | 203.08 |
| Size of the giant component | 1,010 | 1,058 | 1057.1 | 4.86 |
| Density of the network | 0.009% | 0.013% | 0.013% | 0.0002 |

Table 3: Descriptive statistics of the role-specific networks based on the album network.

Consequently, the correlation coefficient of the number of omitted edges and the size of the giant component has been calculated: $r_a = -0.84$. Accordingly, there is a high negative correlation between the number of omitted edges and the size of the giant component. However, the correlation coefficient is not 1, thus, other factors – such as the importance of the omitted edges – might influence the size of the giant component.

Additionally, the impact of particular collaborator roles on the connectivity of the giant component has also been analyzed for each omitted collaborator in the social network. Table 4 lists the descriptive statistics of the role-specific networks based on the social network.

| Characteristic | Min | Max | Mean | $\sigma$ |
|---|---|---|---|---|
| Number of omitted edges | 0 | 23,010 | 836 | 2,316.24 |
| Number of edges | 103,143 | 126,153 | 125,316 | 2,316.24 |
| Size of the giant component | 7,911 | 9,072 | 9,039 | 109.73 |
| Density of the network | 0.002 | 0.003 | 0.003 | 0.0001 |

Table 4: Descriptive statistics of the role-specific networks based on the social network.

The correlation coefficient of the number of omitted edges and the size of the giant component in case of the social network is: $r_s = -0.96$. Accordingly, the correlation coefficient indicates a twelve points higher negative correlation between the number of omitted edges and the size of the giant component than in the album network. This is also underlined by Figure 5 that visualizes the relative size of the giant component in relation to the fraction of omitted links and the corresponding linear regression line for both networks. It can be seen that there is a stronger linear coherence in the social network and that the outliers are not as scattered as in the album network.

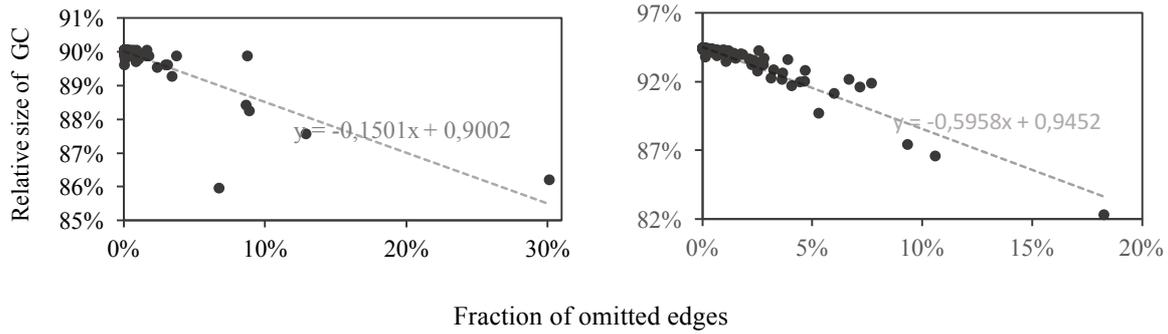

Figure 5: The correlation of the fraction of total links that have been omitted and the relative size of the giant component. The graph on the left represents the album network, whereas the graph on the right represents the social network.

Table 5 list the impact on the connectivity of the giant component of the five most influencing collaborator roles in the album network. Hence, it is seen that the role 'main artist' has the highest impact on the size of the giant component, even though the fraction of omitted links is lower than for other roles. Omitting all edges that are fully dependent on the role 'main artist' decreases the relative size of the giant component by four percentage points from 90% to 86%. On the contrary, the role 'mastered' decreases the size of the giant component only by 3.8 percentage points with a 4.47 higher fraction of omitted links.

| # | Excluded Role | Size of the GC | Fraction of omitted links |
|---|---|---|---|
| 1 | Main Artist | 86.0% | 6.74% |
| 2 | Mastered | 86.2% | 30.11% |
| 3 | Engineer | 87.6% | 12.89% |
| 4 | Photography | 88.2% | 8.88% |
| 5 | Producer | 88.4% | 8.64% |

Table 5: Top 5 of the most influencing collaborator roles in the album network on the size of the giant component.

Accordingly, it is said that the impact of the collaborator roles on the size of the giant component in the album network differs. The impact is not only based on the number of omitted edges, but also on the importance of those edges (i.e. bridges). Hence, the most important collaborator roles in the album network are "main artists" followed by "mastered", "engineer", "photography", and "producer".

Table 6 lists the impact on the connectivity of the giant component of the five most influencing collaborator roles in the social network. In comparison to the album network, the ranking and size of the impact differs: First are the engineers, followed by photography and the main artists. Moreover, it can be seen that the collaborator role with the highest impact also has the highest fraction of omitted links.

| # | Excluded Role | Size of the GC | Fraction of omitted links |
|---|---|---|---|
| 1 | Engineer | 82.37% | 18.24% |
| 2 | Photography | 86.61% | 10.57% |
| 3 | Main Artist | 87.46% | 9.31% |
| 4 | Mastered | 89.72% | 5.26% |
| 5 | Producer | 91.17% | 5.96% |

Table 6: Top 5 of the most influencing collaborator roles in the social network on the size of the giant component.

## 5. Limitations and Future Research

There are some limitations in this paper. This study investigated the influence of particular collaborator roles on the connectivity of the giant component in the specific context of the music industry. Hence, the results of this analysis are probably not applicable to other collaboration contexts, since the collaborations in the music industry may have a specific procedure to form collaborations. Assuming two bands that want to collaborate for an album might not only involve a particular collaborator role but the entire team (e.g. consisting of singers, bassists, drummers, etc.). Thus, further research is necessary to validate and generalize the findings to other collaboration contexts.

Furthermore, since the data for the collaboration network has been selected from two popular lists that claim to name the most important albums of all time, the selected data represents only an extract of all collaborations in the music industry. Hence, the selection process could lead to a selection bias that might influence the results. Accordingly, further research in the context of the music industry should investigate a broader collaboration network by expanding the dataset and validate the results.

Another limitation is the grouping of the collaborator roles. Corresponding to section 3, the roles of the collaborators have been cleaned by creating a more common labeling of the crawled collaborator roles using the defined rules. Nevertheless, out of the 2,216 crawled roles 257 retained. As a result, the grouped roles are still very detailed and could be merged to more generalized roles (for instance, several groups for guitarists exist). Further research could map the roles to more general roles, for example, based on different levels of the production process of an album. Thus, there could be a possible split to managers, musicians, editors, etc.

## 6. Conclusion

An emerging view on collaboration networks has been investigated in this paper: The structure of the giant component has been analyzed regarding the impact of particular collaborator roles. Furthermore, existing research on collaboration networks has been applied to collaboration networks in the music industry.

In general, the existing theories for collaboration networks have been validated: The collaboration network of this paper also constituted a "small world", in which the average distance between two collaborators is small in comparison to the size of the network. Typically, it has been found that the average number of steps to get from one collaborator to a random other one is only four. Moreover, it has been found that the clustering coefficient of the social network is very high with 0.9. This means that two collaborators are much more likely to have collaborated if they have a third mutual collaborator, which has been explained by the existence of typical structures and processes for collaborations in the music industry. These typical structures and processes that lead to a high clustering are also in accordance with existing research on collaboration networks.

Furthermore, the existing research on collaboration networks has been expanded by investigating the impact of particular collaborator roles on the composition of the giant component of both one-mode projections of the collaboration network. It has been found that different roles of collaborators have a different impact on the connectivity of the network and specifically on the giant component of the network. Moreover, it has been found that the impact of a specific collaborator role is not only dependent on the number of connections of the role but also on the importance of these connections.

Finally, it can be said that both one-mode projections can be used to measure the impact of collaborator roles on the connectivity of the network. Each of them has a different focus and leads to different conclusions. In the album network nodes represent teams working on a specific album, whereas the social network represents individual collaborators as nodes. The collaborator roles with the biggest impact on the album network are those that participate in different teams and establish the most bridges between these teams. On the contrary, the collaborator roles with the biggest impact on the social network are those that have a huge number of collaborators and establish the most bridges. In the particular collaboration network of this paper the most influencing roles in the album network were 'main artists' followed by 'mastered' and 'engineered'. In contrast, in the social network the most influencing roles were 'engineer' followed by 'photography' and 'main artist'.